\documentstyle[11pt,epsfig,graphicx,amsmath,amssymb]{article}

\begin{document}
\title{Scaling and a Fokker-Planck model for fluctuations in geomagnetic
indices and comparison with solar wind $\epsilon$ as seen by WIND and ACE}
\author{B. Hnat, S.C. Chapman, G. Rowlands\\
Space and Astrophysics Group, University of Warwick Coventry,
CV4 7AJ, UK}
\date{\today}
\maketitle

\begin{abstract}
The evolution of magnetospheric indices on temporal scales shorter than that
of substorms is characterized by bursty, intermittent events that may arise
from turbulence intrinsic to the magnetosphere or may reflect solar
wind-magnetosphere coupling. This leads to a generic problem of distinguishing
between the features of the system and those of the driver. We quantify scaling
properties of short term (up to few hours) fluctuations in the geomagnetic
indices $AL$ and $AU$ during solar minimum and maximum along with the parameter
$\epsilon$ that is a measure of the solar wind driver. We find that self-similar
statistics provide a good approximation for the observed scaling properties of
fluctuations in the geomagnetic indices, regardless of the solar activity level,
and in the $\epsilon$ parameter at solar maximum. This self-similarity persists
for fluctuations on time scales at least up to about $1-2$ hours. The scaling
exponent of $AU$ index fluctuations show dependence on the solar cycle and the
trend follows that found in the scaling of fluctuations in $\epsilon$. The
values of their corresponding scaling exponents, however, are always distinct.
Fluctuations in the $AL$ index are insensitive to the solar cycle as well as
being distinct from those in the $\epsilon$ parameter. This approximate
self-similar scaling leads to a Fokker-Planck model which, we show, captures
the probability density function of fluctuations and provides a stochastic
dynamical equation (Langevin equation) for time series of the geomagnetic indices.
\end{abstract}

\section{Introduction}
The Earth's magnetosphere can be considered as non-linear, dissipative
system which is driven by the time varying solar wind. Accumulated energy is
ultimately dissipated, at least in part, through a system of currents generated
in the auroral zones of the ionosphere. These currents produce small changes in
the terrestrial magnetic field which can be used to monitor magnetospheric
activity. The complex behavior of the magnetosphere, as suggested by many
observations (see, for example, \cite{horton,lewis,sitnov,takalo00,vassiliadis00,voros02}),
could then be attributed either to intrinsic magnetospheric processes, the
complex nature of its coupling with the solar wind and the ionosphere or both.

Recent observations suggest that the multi-scale nature of this coupling
is a fundamental aspect of magnetospheric dynamics (see, for example
\cite{chang92,chapman01,klimas,ukhorskiy,vassiliadis03,weigel03b}). Evidence is
provided by a variety of observations which exhibit statistical properties
previously identified as hallmarks of multi-scale systems. For example, bursty
transport events have been reported in the magnetotail\cite{angelopoulos92}
and their auroral signatures suggest self-similar statistics
\cite{lui,uritsky01}.
\begin{figure}
\epsfsize=0.5\textwidth \centerline{\leavevmode\epsffile{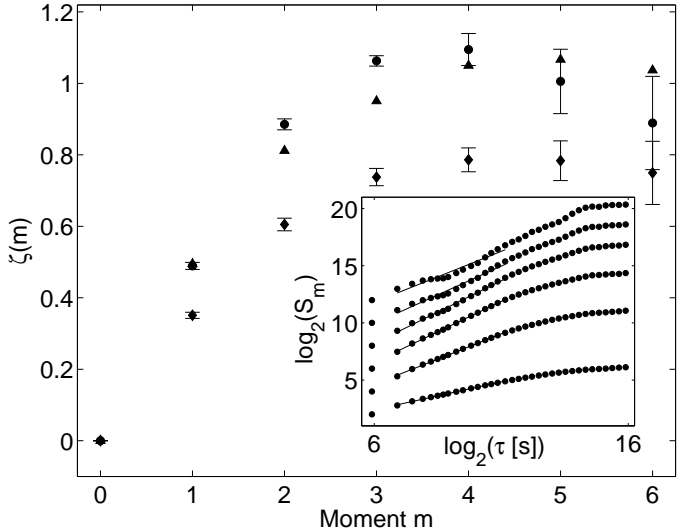}}
\caption{Exponents of unconditioned generalized structure functions as a
function of order $m$ for fluctuations in the $\epsilon$($\diamond$),
$AU$($\circ$) and $AL$($\triangle$) index during solar maximum. The inset
shows structure functions $S_m$ of orders $m=1-4$ for the $\epsilon$
parameter.}
\label{fig1}
\end{figure}
The fluctuations in the ground based measurements of the magnetic field are
non-Gaussian and also exhibit scaling \cite{consolini96,kovacs,voros98}. In the
context of time series analysis, geomagnetic indices are of particular interest
as they provide a global measure of magnetospheric output and are evenly sampled
over a long time interval. These indices also show non-Gaussian statistics of
fluctuations and anomalous scaling over the short time scales of up to few hours
\cite{consolini98,takalo93,hnat03a,stepanova,tsurutani}.
The extent to which observed statistical features of the geomagnetic
indices arise directly from those of the solar wind driver or the auroral
currents is of fundamental interest. This is an example of the generic problem
of distinguishing between features intrinsic to a driven system and those
in the driver, when both show scaling. Some recent studies has focused on
direct comparison of scaling properties of the driver with these found in the
geomagnetic indices\cite{freeman,uritsky01,hnat03a} to establish whether, to
the lowest order, they are directly related. 

The difficulty with interpreting these observations arises from the fact
that statistical features described above can be recovered from many existing
models. Self-Organized Criticality (SOC) and turbulence have both been
extensively used \cite{angelopoulos99,consolini98},\\ \cite{kozelov,uritsky98} in
the past. Practically, one needs to obtain experimental constrains with which
different models with similar characteristics can be tested. In this paper we
present one possible approach to characterizing the time series in the context
of scaling that does not rely on a specific model of multi-scale
systems\cite{sornette,hnat03a}. The aim is to obtain statistical scaling
properties directly from the data.

Here, we will examine the statistical properties of Akasofu's
$\epsilon$\cite{akasofu} parameter, which represents the energy input from the
solar wind into the magnetosphere, and that of magnetospheric response as seen
by the geomagnetic indices.
Previously, scaling has been quantified over a $10$ year data set for the
indices and a comparison between $\delta \epsilon$ and the indices included,
but was not restricted to, the solar minimum ($1984-1987$)\cite{hnat03a}.
Here, we will perform this comparison over intervals of solar minimum and
maximum separately. The statistical description of the experimental data will
be extended to $10$ standard deviations of the fluctuations.

To facilitate the comparison of all considered quantities we will first explore
to what extend their fluctuations exhibit approximate self-similar scaling for
temporal scales of $1-2$ hours.
The quality of this self-similar approximation combined with values of the
scaling exponents obtained at the solar minimum and maximum can be used to
characterize each quantity. We will see that values of scaling exponents on
these temporal scales for the geomagnetic indices are different from these
found in the solar wind $\epsilon$ regardless of the phase of the solar cycle.
Remarkably, the scaling exponent of the $AL$ index is unchanged between solar
minimum and maximum whereas the $AU$ scaling exponent changes with the solar
cycle.
\begin{figure}
\epsfsize=0.5\textwidth \centerline{\leavevmode\epsffile{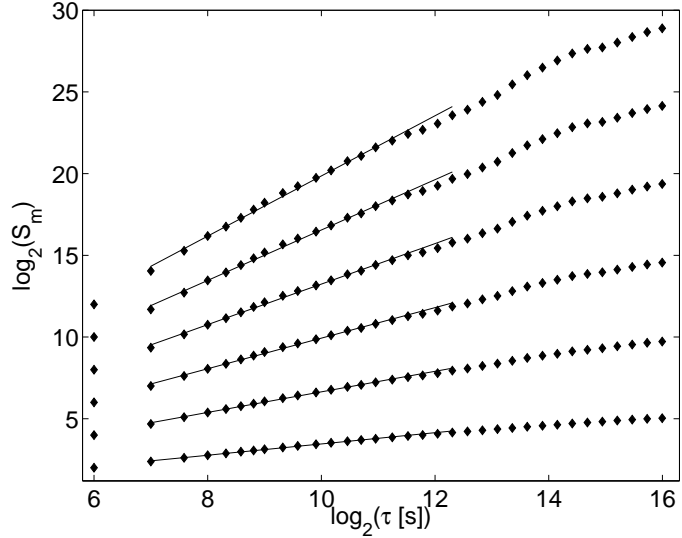}}
\caption{Structure functions $S_m$ of orders $m=1-6$ for fluctuations in
the $\epsilon$ parameter at solar maximum.}
\label{fig2}
\end{figure}
In this respect, the $AU$ index seems to follow the trend found in the driver,
$\epsilon$ i.e., the value of scaling exponent increases with increasing solar
activity. We then construct a Fokker-Planck model for fluctuations in the
geomagnetic indices and the $\epsilon$ at solar maximum as these exhibit the
most satisfactory self-similar scaling.
\begin{figure}
\epsfsize=0.5\textwidth \centerline{\leavevmode\epsffile{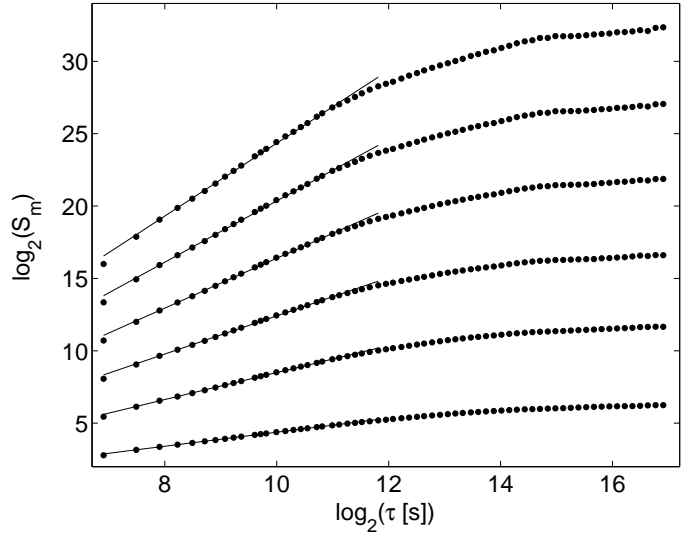}}
\caption{Same as figure \ref{fig2} for the $AU$ index.}
\label{fig3}
\end{figure}
This allows us to obtain analytically a functional form of the fluctuation
probability density function (PDF) which we can then check against the data.
A stochastic dynamical model can then be formulated by considering the most
general form of the Langevin equation and deriving functional forms of the
coefficients that are consistent with the Fokker-Planck equation (see, for
example \cite{hnat03b}).

\section{Data and Methods}
\subsection{Data sets}
To facilitate this analysis we used multiple data sets that spanned over
different phases of the solar cycle. Two year intervals of data were selected
centered on solar minimum and solar maximum. The solar wind data
were obtained from WIND and ACE spacecraft observations. These were collected
in the vicinity of the L1 point approximately $1$ AU (Astronomical Units) from
the Sun.
\begin{table}[H]
\begin{center}
\caption{Data sets description}
\begin{tabular}{cccccc}
\hline
\hline
Quantity	&Solar cycle	& dt[s]	& Dates	&N[mln]	&Source\\
\hline 
$AL$, $AU$&Min&$60$&$01/85-12/86$&$1.05$&WDC STP\\
$AL$, $AU$&Max&$60$&$01/79-12/80$&$1.05$&WDC STP\\
$\epsilon$&Min&$92$&$08/95-07/97$&$0.63$&WIND\\
$\epsilon$&Max&$64$&$01/00-12/01$&$0.68$&ACE\\
\hline
\hline
\end{tabular}
\label{tab1}
\end{center}
\end{table}
The periods of coverage, final sampling frequencies and number of samples
are given in the Table 1. The geomagnetic indices and the corresponding
spacecraft data sets are not contiguous. The calibrated geomagnetic data set,
from which intervals of interest has been selected, spans from January $1978$
to December $1988$ inclusive, while the spacecraft data are available starting
from $1995$ for WIND and $1998$ for ACE. This available data coverage does
not permit examination of successive solar cycles. We thus need to assume that
the statistical properties of fluctuations are invariant from one solar cycle
to the next. In the case of the spacecraft data, these include slow and fast solar
wind streams. 

The solar wind velocity measurements, provided by the SWE instrument on board
of WIND and ACE spacecraft, have varying temporal resolutions. In the case of
WIND this resolution is in the range of $75-98$ seconds while for the ACE
spacecraft it changes between $60-120$ seconds. The magnetometer data sets,
on the other hand, have fixed temporal resolution of $46$ seconds for WIND MFI
instrument and $16$ seconds in the case of the ACE magnetometer. The SWE data
sets have been then re-sampled using linear interpolation to give uniform
resolution of $92$ seconds for WIND (twice the magnetometer resolution) and
$64$ seconds for the ACE spacecraft (four times magnetometer resolution).
No other post processing, such as detrending or smoothing, was applied to data.
The $\epsilon$ parameter is defined\cite{akasofu} in SI units (Watts) as
$\epsilon=v(B^2/\mu_0) l_0^2 \sin^4(\Theta /2)$, where $l_0\approx 7R_E$ and
$\Theta=\arctan(|B_y|/B_z)$, and was calculated from WIND and ACE spacecraft
key parameter databases.
\begin{figure}
\epsfsize=0.5\textwidth \centerline{\leavevmode\epsffile{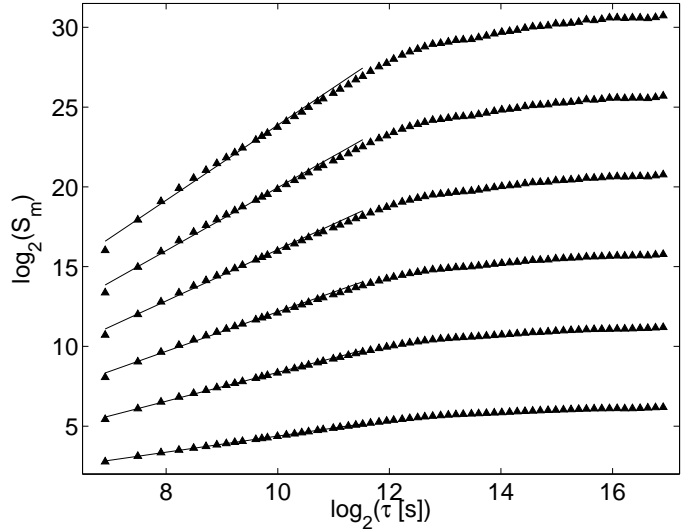}}
\caption{Same as figure \ref{fig2} for the $AL$ index.}
\label{fig4}
\end{figure}

All techniques discussed here are based on differencing of the original time
series over a range of temporal scales $\tau$. This method is often used
in turbulence studies to compare the properties of fluctuations on different
spatio-temporal scales (see, for example, \cite{frisch}). For a given time
series $x(t)$ a set of differences $\delta x(t,\tau)=x(t+\tau)-x(t)$ will then
capture fluctuations on temporal scale $\tau$. Here, we will examine the
statistical properties of the PDF of fluctuations $\delta x(t,\tau)$. The
$\tau$ parameter will be given in power law form such as
$\tau=\delta t_{AU} (1.2)^n$ seconds, where $\delta t_{AU}$ is a sampling time
of the $AU$ time series (here, $1$ minute) and $n\geq1$ is an integer.
This choice of $\tau$ gives a uniform distance between points when plotted on
the logarithmic scale while the small base of the power law ($1.2$) assures
that the adequate number of temporal scales are explored.
We stress that the differencing is performed only if both $x(t+\tau)$ and
$x(t)$ exist and are separated by time interval $\tau$.
              
\subsection{Statistical Methods}
Generalized structure functions (GSF) $S_m$ are widely used to characterize
non-Gaussian processes\\ \cite{frisch,hnat03a}.
\begin{figure}
\epsfsize=0.5\textwidth \centerline{\leavevmode\epsffile{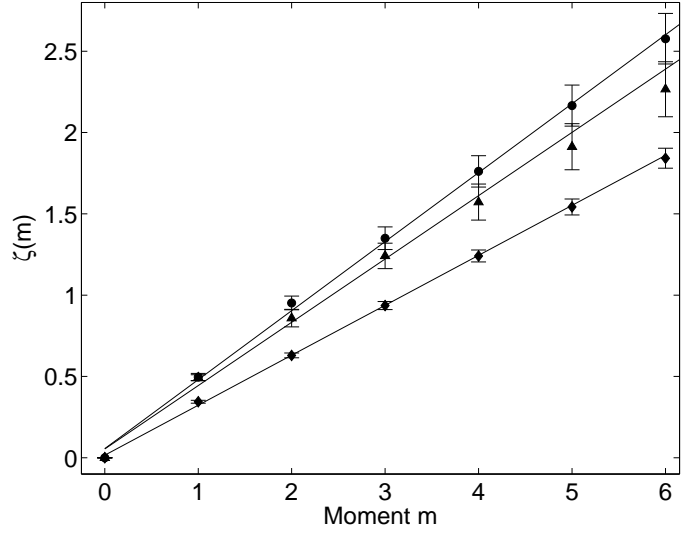}}
\caption{Exponents of conditioned generalized structure functions as a
function of order $m$ for fluctuations in the $\epsilon$($\diamond$),
$AU$($\circ$) and $AL$($\triangle$) index during the solar maximum.}
\label{fig5}
\end{figure}
These functions can be defined for fluctuations $\delta x(t,\tau)$ as
$S_m(\tau)\equiv\langle|\delta x|^m\rangle$, where $m$ can be any real number,
not necessarily positive. If $S_m$ exhibits scaling with respect to the time
lag $\tau$ we also have $S_m \propto \tau^{\zeta(m)}$. In this case a log-log
plot of $S_m$ versus $\tau$ should reveal a straight line for each $m$ and the
gradients correspond to values of $\zeta(m)$. Generally, $\zeta(m)$ can be a
non-linear function of order $m$, however if $\zeta(m)=\alpha m$ ($\alpha$
constant) then the time series is self-similar (or more precisely, self-affine)
with single scaling exponent $\alpha$. This special case leads immediately to a
Fokker-Planck description \cite{hnat03b}. The difficulty with computing GSF for
higher orders, say, $m>4$ arises from the slow convergence of this method and
its sensitivity to large statistical errors in extremal events in the tails of
the distribution. These effects can, as we shall see, lead to large errors in
the estimation of $\zeta(m)$ (see also \cite{horbury} for the discussion of
error estimation for structure functions). One possible approach is to eliminate
these extreme events from the fluctuation time series $\delta x(t,\tau)$ in a
way that is consistent with the growth of the self-similar fluctuations' range
on each temporal scale. This process is referred to as conditioning. Previously,
a similar technique based on the wavelet filters has been used to separate the
intermittent parts of the signal from the homogeneous noise in the $AE$ index
data\cite{kovacs}. We will condition our GSFs by imposing a threshold $A$ on
the fluctuation size\cite{hnat03a}. The threshold will be based on the
standard deviation of the differenced time series for a given $\tau$,
$A(\tau)=10\sigma(\tau)$. Under conditioning, the GSF can be expressed in term
of the fluctuation PDF as:
\begin{equation}
<|\delta x|^m>=\int_{-A}^A |\delta x|^m P(\delta x,\tau)d(\delta x).
\label{gsfcond}
\end{equation}
This procedure is then consistent with scaling $\zeta(m)=m \alpha$ if it is
present in the data, but for threshold $A$ sufficiently large it does not
enforce it on the data.
\begin{figure}
\epsfsize=0.5\textwidth \centerline{\leavevmode\epsffile{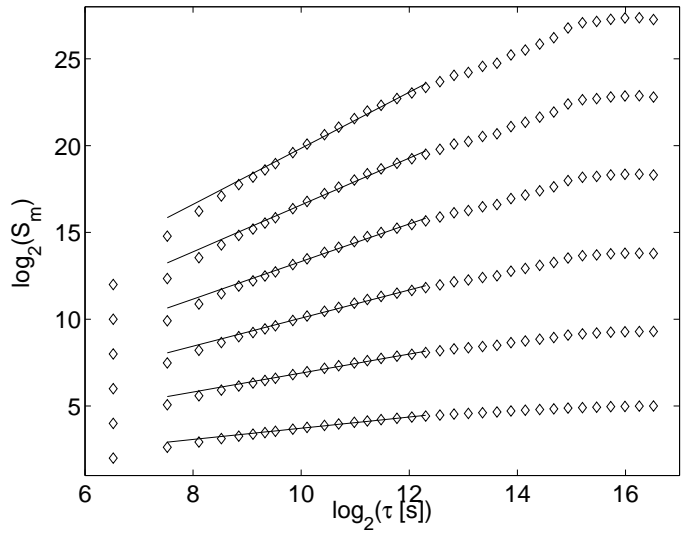}}
\caption{Structure functions $S_m$ of orders $m=1-6$ for fluctuations in
the $\epsilon$ parameter at solar minimum.}
\label{fig6}
\end{figure}

The PDF rescaling technique is a generic and model independent method of
testing for statistical self-similarity in the data set. If the data is
self-similar, then a single argument representation of the fluctuation PDF,
$P_s(\delta x_s)$, can be found in the form:
\begin{equation}
P(\delta x,\tau)=\tau^{-\alpha} P_s(\delta x \tau^{-\alpha}),
\label{rescl}
\end{equation}
where $\alpha$ is the rescaling exponent. Substituting the rescaled quantities
$P_s$ and $\delta x_s=\delta x \tau^{-\alpha}$ into the GSF definition given by
(\ref{gsfcond}) we obtain:
\begin{equation}
<|\delta x|^m>=\tau^{m\alpha} \int_{-A_s}^{A_s} |\delta x_s|^m P(\delta
x_s)d(\delta x_s) \propto \tau^{\zeta(m)},
\label{gsfrecl}
\end{equation}
where the integral now has no explicit dependence on temporal scale $\tau$.
This then immediately relates the PDF rescaling to self-similar scaling of
the GSF with $\zeta(m)=m\alpha$.

In this approach PDFs are generated using non-overlapping intervals of the
original data, i.e., $\delta x(t,\tau)=x[m\tau]-x[(m-1)\tau]$. The method
assures that fluctuations are not temporally correlated--an important
assumption for a Fokker-Planck model we will consider later.
These two methods are thus complementary as one provides a scaling exponent
while the other gives an underlying probability distribution of fluctuations.

\section{Results and discussion}
\subsection{GSF analysis}
We will first present scaling properties of the GSFs for the indices and the
$\epsilon$ parameter during solar minimum and solar maximum. To illustrate the
effect of conditioning, we first show, in the inset of figure \ref{fig1}, a
log-log plot of structure functions $S_m$ obtained for fluctuations in the raw time
series of the $AU$ index at solar maximum for orders $1 \leq m \leq 6$. We see that,
for orders $m>3$ there is no clear evidence of scaling--the points do not lie on
straight lines.
\begin{figure}
\epsfsize=0.5\textwidth \centerline{\leavevmode\epsffile{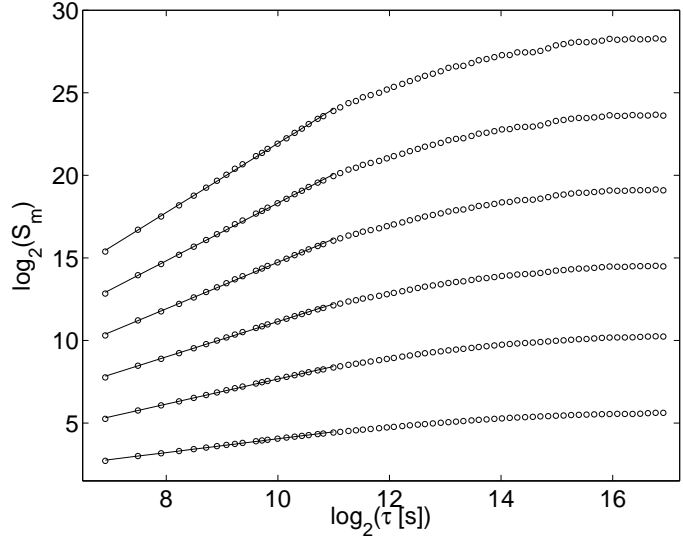}}
\caption{Same as figure \ref{fig6} for the $AU$ index.}
\label{fig7}
\end{figure}
Similar lack of scaling was also found for fluctuations in the $AL$ index and the
$\epsilon$ parameter. The main panel of figure \ref{fig1} shows exponents $\zeta(m)$
obtained by performing linear fits to logarithms of moments $log[S_m(log(\tau))]$.
We see that the curves $\zeta(m)$ are not monotonic functions of $m$, excluding the
possibility of multi-fractal scaling.
We now condition this data set as discussed above, to check if true scaling properties
are not obscured by poor statistics of extreme and very rare events.

Figures \ref{fig2}-\ref{fig4} show a log-log plot of structure functions $S_m$
for the $\delta \epsilon$, $\delta (AU)$ and $\delta (AL)$ indices at solar maximum
and for order $m$ from $1$ to $6$. The main indication of successfully recovered
scaling after the conditioning process is the quality of the linear fit to
$log[S_m(\tau)]$ versus $log(\tau)$.
\begin{figure}
\epsfsize=0.5\textwidth \centerline{\leavevmode\epsffile{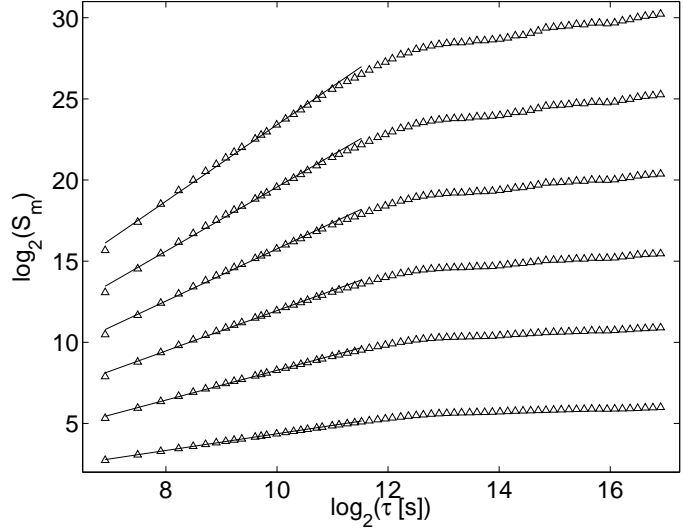}}
\caption{Same as figure \ref{fig6} for the $AL$ index.}
\label{fig8}
\end{figure}
We clearly recover a family of straight lines with slopes $\zeta(m)$, up to order
$m=6$ for fluctuations in $AU$ and $AL$ indices and in $\epsilon$. This scaling
extends up to temporal scales of $\sim1$ to $\sim2$ hours in good agreement with
these reported earlier\cite{takalo93,tsurutani,hnat03a}. Figure \ref{fig5} shows
that fluctuations in all quantities, at solar maximum, exhibit approximate
self-similar scaling to within statistical errors. The size of these error bars
combined with the convex shape of the function $\zeta(m)$ for the $AL$ index also
allows a weakly multi-fractal interpretation of the scaling. We stress, however, that
the error bars in figure \ref{fig5} do not include many other uncertainties (not
statistical) that are difficult to estimate. For example, the WIND spacecraft
magnetometer data has absolute accuracy of about $0.1$nT and the indices data
have integer values (also in units of nT). Such discreteness in the time series
may lead to erroneous estimates of low order moments while the finite size of
the data could alter true scaling of the high order moments. Independent of
any given choice of a model for the functional form of $\zeta(m)$ we can
perform a direct comparison between the $\zeta(m)$ measured for the different
quantities at solar minimum and maximum. In order to develop a Fokker-Planck approach
we will then make a further step and assume that a reasonable approximation is given
by $\zeta(m)=m\alpha$, that is, self-similar scaling.

Figures \ref{fig6}-\ref{fig8} show structure functions $S_m$ for all quantities at
solar minimum and with order $m$ varying again from $1$ to $6$. We see that
moments of fluctuations for the geomagnetic indices show satisfactory scaling up to
temporal scales of $1-2$ hours. In the case of $\epsilon$ at solar minimum there is
a departure from a single set of scaling exponents $\zeta(m)$ for the smallest
time scales $\tau<12$ minutes. To facilitate a comparison with conditions at solar
maximum and with the indices we will fit straight lines to obtain $\zeta(m)$ for
$\tau=[12,90]$ minutes bearing in mind that this does not capture the behavior of
fluctuations on the smallest time scales.
\begin{figure}
\epsfsize=0.5\textwidth \centerline{\leavevmode\epsffile{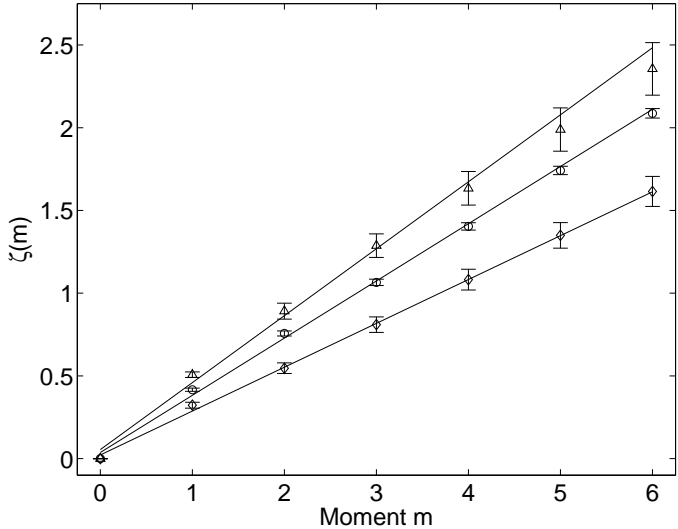}}
\caption{Same format as figure \ref{fig5} for data around solar minimum.}
\label{fig9}
\end{figure}
This change in scaling properties for $\epsilon$ may reflect differences between
solar wind evolution at solar minimum and maximum related to physical properties of
slow and fast wind components\cite{pagel01}. Figure \ref{fig9} is constructed
identically to figure \ref{fig5} and shows scaling exponents $\zeta(m)$ for solar
minimum. 

\begin{figure}
\epsfsize=0.75\textwidth \centerline{\leavevmode\epsffile{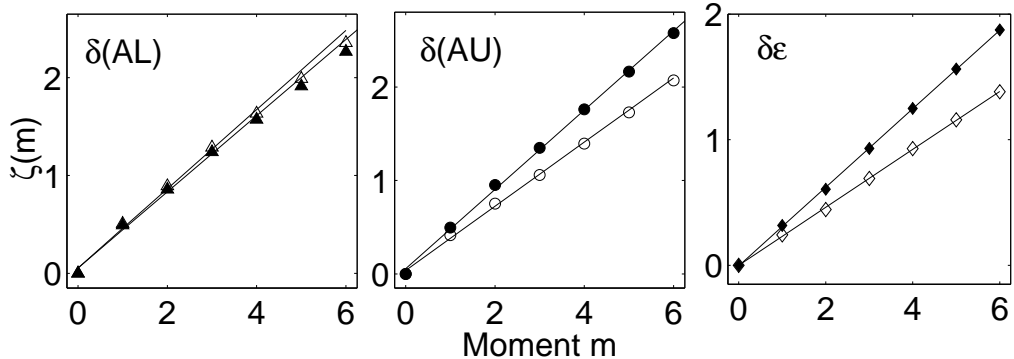}}
\caption{Comparison of functions $\zeta(m)$ during solar minimum and maximum
for fluctuations in all three quantities: (a)-$AL$ index, (b)-$AU$ index and
(c)-$\epsilon$.}
\label{fig10}
\end{figure}

To make a comparison between behavior at maximum and minimum we plot, in figure
\ref{fig10}, exponents $\zeta(m)$ at solar minimum and maximum overplotted for
$AL$, $AU$ and $\epsilon$ respectively.
Examining these figures we conclude that the scaling properties of the $AL$
index fluctuations are remarkably insensitive to the change in solar activity.
\begin{table}[H]
\begin{center}
\caption{Scaling indices derived from GSF analysis}
\begin{tabular}{cccc}
\hline
\hline
Quantity	&Solar cycle	& $\alpha$ from GSF	& $\tau_{max} [hr]$\\
\hline 
$\delta AU$&Min&$-0.35\pm0.01$&$\sim1$\\
$\delta AU$&Max&$-0.43\pm0.01$&$\sim1$\\
$\delta AL$&Min&$-0.39\pm0.02$&$\sim2$\\
$\delta AL$&Max&$-0.37\pm0.03$&$\sim2$\\
$\delta \epsilon$&Min&$-0.26\pm0.02$&$\sim2$\\
$\delta \epsilon$&Max&$-0.32\pm0.02$&$\sim2$\\
\hline
\hline
\end{tabular}
\label{tab2}
\end{center}
\end{table}
The values of $\zeta(m)$ and the corresponding scaling exponents are the same,
to within the statistical error for solar minimum and maximum. On the other
hand the scaling exponents of fluctuations in both $\epsilon$ and the $AU$ index
vary with the solar cycle. The scaling of $\delta (AU)$ is distinct from these
of $\delta (AL)$ and $\delta \epsilon$ but follows the trend of
$\delta \epsilon$. A possible interpretation of these observations is that the
$AL$ index fluctuations more closely reflect the internal dynamics of the magnetotail
and are insulated from solar cycle related changes in the solar wind. In contrast,
the $AU$ index is more strongly coupled to solar cycle associated changes in
the solar wind driver. This is consistent with our understanding of the global
roles of these indices (eg., \cite{baumjohann}). The fluctuations in $AU$,
however, have values of scaling exponents different from that observed for the
driver $\epsilon$ at solar minimum and maximum, which may suggest that
(i) $\epsilon$ does not completely capture all relevant information about the
driver, (ii) the indices do not fully capture the magnetospheric response or
(iii) the difference reflects the non-linear nature of the solar
wind-magnetosphere coupling.

If we compare the scaling exponents of $\delta (AU)$ and $\delta \epsilon$
during solar minimum and maximum we see that both quantities follow a similar
trend. The exponents have values closer to that of Brownian motion ($0.5$) during
solar maximum as compared to minimum. Closer examination of scaling exponents
for fluctuations in the $\epsilon$ and the $AU$ index reveals that the
difference $\alpha^{AU}-\alpha^{\epsilon} \approx 0.06$ is almost identical for
solar minimum and solar maximum period, to within the statistical error. This
could indicate that the ``conversion rate" of fluctuations in the driver to
those in the $AU$ index is nearly constant and independent of the strength of
the driver.

All scaling exponents $\alpha$ derived by fitting $\zeta(m)=m\alpha$ are given
in the Table 2 together with the approximate maximum temporal scale
$\tau_{max}$ for which self-similarity can be identified in the differenced
time series. These temporal scales have been derived using $R^2$ goodness of
fit analysis for moment with $m=2$. We have also verified that GSF analysis of
combined time series over solar minimum and maximum recovers results presented
in our previous work\cite{hnat03a}.

\subsection{Probability Density Function (PDF) Rescaling}
We now present the results of the PDF rescaling analysis which allows us to
compare directly the PDFs of the studied parameters at solar minimum and maximum.
Due to the rather poor scaling of the higher moments for the $\epsilon$
fluctuations at solar minimum we will not apply this rescaling to their PDFs.
\begin{figure}
\epsfsize=0.5\textwidth \centerline{\leavevmode\epsffile{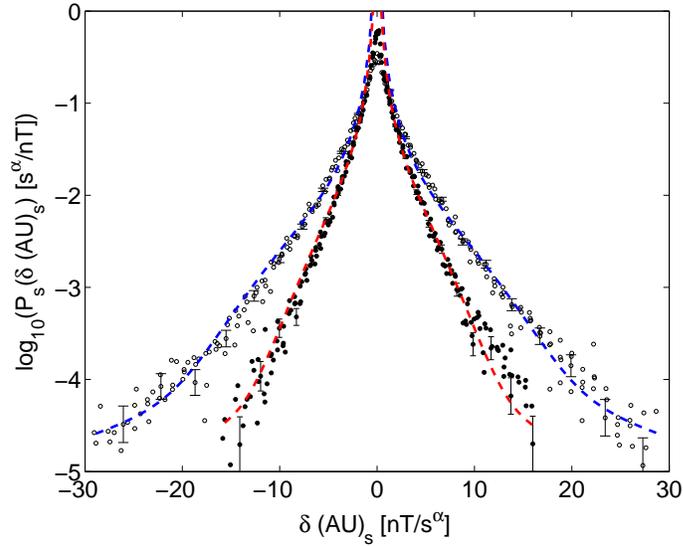}}
\caption{Rescaled PDFs of the $\delta (AU)$ during solar minimum
(empty symbols) and maximum (filled symbols). Symbols correspond to
temporal scales $\tau=10$, $16$, $26$ and $42$ minutes. The dashed line
represents a solution of the Fokker-Planck model (\ref{mdl2}) with parameters
given in Table 3.}
\label{fig11}
\end{figure}
We simply state that our previous work suggested that PDFs of fluctuations in the
geomagnetic indices and that of the $\epsilon$ fluctuations differed
significantly when considered time interval spanned more then a solar
minimum \cite{hnat03a}.

\begin{figure}
\epsfsize=0.5\textwidth \centerline{\leavevmode\epsffile{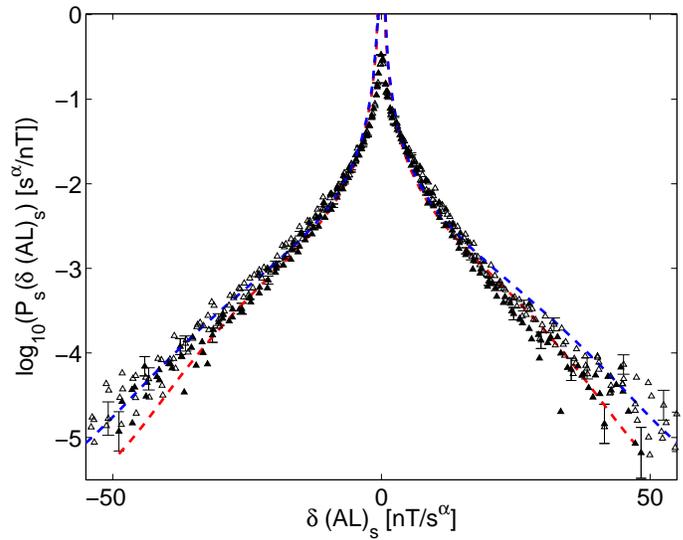}}
\caption{Same format as figure \ref{fig11} for the $AL$ index fluctuations.}
\label{fig12}
\end{figure}

Figures \ref{fig11} and \ref{fig12} show rescaled PDFs of fluctuations in the
$AU$, $AL$ index respectively for solar minimum (empty symbols) and solar
maximum (filled symbols) while figure \ref{fig13} shows these PDFs for the
$\epsilon$ parameter but only at solar maximum.
These PDFs correspond to the function $P_s(\delta x_s)$ in equation
(\ref{rescl}). Each plot shows overplotted $P_s(\delta x_s)$ at four temporal
scales, $\tau=10$, $16$, $26$ and $42$ minutes. 
These figures show data up to $10$ standard deviation on any  given temporal
scale--consistent with the conditioning procedure described above. All rescaling
exponents $\alpha$ used to construct these plots, are taken directly from the
GSF analysis. We find that, when solar minimum and maximum data sets are taken
separately, these PDFs collapse on a single curve after rescaling (\ref{rescl})
is applied. The quality of the collapse for the PDFs was checked
using the Smirnov-Kolmogorov\cite{numrecep} test and the significance of the
null hypothesis (both curves drawn from the same distribution) was always found
to be above the $0.975\pm0.05$ level.

The rescaling confirms what we have already found by applying GSF analysis,
in that a single exponent $\alpha$ is sufficient to give close correspondence
of the curves. As we have shown in equation (\ref{gsfrecl}) this is consistent
with approximate self-similar scaling $\zeta(m)=m\alpha$ from the GSF analysis.
Once rescaled, using the values of exponents obtained separately for solar
minimum and maximum, we see that the curves are distinct and the difference
is most clear for the $AU$ index shown in figure \ref{fig11}.

We also compared the functional form of these curves by applying normalization
to their respective standard deviation on a given temporal scale, $\sigma_s(\tau)$.
We found that the normalized PDFs for maximum and minimum are indistinguishable
within the errors for $AU$ and $AL$. Similar results were reported for ground based
measurements of the magnetic field \cite{weigel03a} where authors also found that
the statistics of fluctuations, when normalized to the standard deviation, is not
sensitive to changing solar wind conditions.

\section{The Fokker-Planck Model}
The Fokker-Planck (F-P) equation provides an important link between statistical
properties of the system and the dynamical approach expressed by the Langevin
equation\cite{kampen}.
The F-P approach can be readily applied if fluctuations are self-similar and
statistically independent (uncorrelated)\\ \cite{kampen}.
The above analysis suggests that self-similar scaling is a reasonable
approximation to the data. The independent nature of increments is enforced
by considering non-overlapping intervals for differencing, as discussed in
Section II B.

In the most general form the F-P equation can be written as:
\begin{equation}
\frac{\partial{P}}{\partial{\tau}}=
\nabla_{\delta x} (A(\delta x)P + B(\delta x)\nabla_{\delta x}P), 
\label{f-p}
\end{equation}
where $P \equiv P(\delta x,\tau)$ is a PDF for the differenced quantity
$\delta x$ that varies with time $\tau$ and $A(\delta x)$ and $B(\delta x)$ are
transport coefficients which vary with $\delta x$. It can be shown that, under
the assumption of power law scaling $A(\delta x) \propto \delta x^{1-1/\alpha}$
and $B(\delta x) \propto \delta x^{2-1/\alpha}$, a class of self-similar
solutions of (\ref{f-p}) can be found that also satisfies the rescaling
relation (\ref{rescl})\cite{hnat03b}. These assumptions combined with the use
of rescaled variables $\delta x_s=\delta x \tau^{\alpha}$ and $P_s$ lead to the
following equation:
\begin{equation}
\frac{b_0}{a_0}(\delta x_s)\frac{dP_s}{d(\delta x_s)}+P_s+\frac{\alpha}{a_0}
(\delta x_s)^{\frac{1}{\alpha}}P_s = C,
\label{mdl1.5}
\end{equation}
where $a_0$, $b_0$ , $C$ are constants and $\alpha$ is the rescaling exponent
derived, for example, from GSF analysis. The general solution of (\ref{mdl1.5})
is given by the sum of homogeneous and inhomogeneous solutions\cite{hnat03b}:
\begin{eqnarray}
P_s(\delta x_s)=\frac{a_0}{b_0}\frac{C}{|\delta x_s|^{a_0/b_0}}
exp\left(-\frac{\alpha^2}{b_0}(|\delta x_s|)^{1/\alpha}\right)\nonumber \\ \times \int_0^{\delta x_s} \frac{exp\left(\frac{\alpha^2}{b_0}(\delta x_s')^{1/\alpha}\right)}{(\delta x_s')^{1-a_0/b_0}}d(\delta x_s') + k_0H(\delta x_s),
\label{mdl2}
\end{eqnarray}
where $k_0$ is a constant and $H(\delta x_s)$ is the homogeneous solution:
\begin{equation}
H(\delta x_s)=\frac{1}{|\delta x_s|^{a_0/b_0}}
exp\left(-\frac{\alpha^2}{b_0}(|\delta x_s|)^{1/\alpha}\right).
\label{homsol}
\end{equation}
The simple model described above assumes that self-similar scaling persists
for all $\delta x$. This assumption is expected to hold for a physical system
for a large but finite range of $\delta x$. In particular, it will break down as
$\delta x_s \rightarrow 0$ giving a singularity in the solution $P_s$ as
$\delta x_s \rightarrow 0$.This singularity, however, is integrable so that
$\int_{-\infty}^{\infty} P_s d(\delta x_s)$ is finite. 

\begin{figure}
\epsfsize=0.5\textwidth \centerline{\leavevmode\epsffile{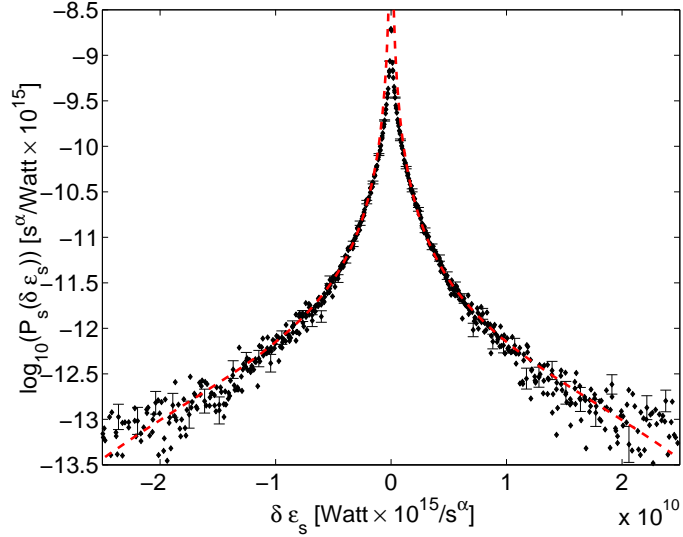}}
\caption{Same format as figure \ref{fig11} for the $\epsilon$ parameter
fluctuations at solar maximum.}
\label{fig13}
\end{figure}

We have found that fluctuations in the geomagnetic indices in solar minimum and
maximum and these in $\epsilon$ at solar maximum exhibit self-similar statistics
to a reasonable approximation. We will now show that the functional form of the
PDF obtained from the F-P model (\ref{mdl2}) is a good approximation for the
observed rescaled distribution $P_s(\delta x_s)$ of fluctuations shown in
figures \ref{fig11}-\ref{fig13}.
On figures \ref{fig11}-\ref{fig13} we have overplotted solution (\ref{mdl2}),
shown by thick dashed line, with $\alpha$ taken to be that obtained from the GSF
analysis. Table 3 gives values of all parameters assumed for each of the
plotted solutions. We see that all PDFs shown in figures \ref{fig11}-\ref{fig13}
are well approximated by their F-P solutions. In the case of the geomagnetic
indices some departures of the predicted curves from the observed distributions
do occur and can be attributed to the asymmetry of these observed PDFs.

We note an obvious departure of our predicted curves from the measured PDF for
the smallest fluctuations, in all considered cases. This is due to the
functional form of (\ref{homsol}) where $H(\delta x_s) \rightarrow \infty$ when
$\delta x_s \rightarrow 0$ arising from the assumption that the self-similar
scaling extends to arbitrary small fluctuations. To model this part of the
curve, we would need to include the scaling, or lack of thereof, introduced by
the uncertainty in the measurements. We would expect such processes
to be dominant for the smallest fluctuations. For example, if we assume that
the smallest fluctuations are dominated by Normally distributed noise, then a
diffusion model with a constant diffusion coefficient $D_0$ could, in principle,
be used to tame this singular behavior.
\begin{table}[H]
\begin{center}
\caption{Values of parameters used for F-P model solutions plotted in figures
\ref{fig11}-\ref{fig13}.}
\begin{tabular}{cccccc}
\hline
\hline
Quantity	&Solar cycle	&$b_0$	&$b_0/a_0$	& $k_0$	& C $[\times
10^{-5}]$\\
\hline 
$\delta AU$&Min&$170 $&$1.875$&$0.28$&$32.5$\\
$\delta AU$&Max&$16  $&$1.875$&$0.20$&$26.4$\\
$\delta AL$&Min&$2200$&$1.8$&$0.36$&$7.77$\\
$\delta AL$&Max&$1000$&$1.8$&$0.32$&$6.66$\\
$\delta \epsilon$&Max&$4\times10^{12}$&$2.15$&$5.3\times10^{10}$&$2.84\times
10^{-10}$\\
\hline
\hline
\end{tabular}
\label{tab3}
\end{center}
\end{table}
This stochastic approach can be extended to obtain the Langevin equation for
the dynamics of the fluctuations\cite{hnat03b}. The Langevin equation can be
written in the most general form as:
\begin{equation}
\frac{d(\delta x)}{dt}=\beta(\delta x)+\gamma(\delta x)\xi(t),
\label{langevin}
\end{equation}
where the random variable $\xi(t)$ is assumed to be $\delta$-correlated.
Equation \ref{langevin} can be transform into purely additive noise form:
\begin{equation}
\frac{dz}{dt} = \frac{\beta(z)}{\gamma(z)} + \xi(t),
\label{langevin1}
\end{equation}
where $z=\int_0^{\delta x} 1/\gamma(\delta x') d(\delta x')$.
It has been shown \cite{hnat03b} that one can obtain a functional form of
coefficients $\beta(\delta x)$ and $\gamma(\delta x)$ in terms of $a_0$,
$b_0$ (from equation \ref{mdl1.5}) and the scaling exponent $\alpha$.
Such an equation provides a dynamical model for time series with the required
statistical properties.

\section{Summary}
The response of the Earth's magnetosphere to the solar cycle and, by
implication, a changing character of solar wind activity, illuminates the
interplay between intrinsic magnetospheric dynamics and solar
wind-magnetosphere coupling.
Statistical studies provide a simple and yet unifying way to quantify this
behavior in the context of models for intermittency. In this paper we
considered scaling properties of the solar wind driver, quantified by the
$\epsilon$ parameter, and geomagnetic indices during solar minimum and maximum.
We find that:
\begin{enumerate}
\item{Fluctuations in the geomagnetic indices show approximate statistical
self-similarity for a range of temporal scales. Fluctuations in the $\epsilon$
at solar minimum show departure from scaling for $\tau < \sim10$ minutes. The
self-similar scaling emerges as a reasonable approximation for fluctuations
$\delta \epsilon$ at solar maximum. Fluctuations in the geomagnetic indices
exhibit self-similar scaling on temporal scales between $\sim2$ minutes to
$\sim 1-2$ hours. The fluctuations in $\epsilon$ scales from $\sim2$ minutes
to $\sim1.5$ hours, but only at solar maximum.}
\item{Fluctuations in the $AL$ index exhibit scaling properties insensitive to the
phase of the solar cycle.}
\item{The scaling exponent of $\delta(AU)$ changes with the solar cycle and
the trend follows that of the $\epsilon$ parameter}
\item{The value of the scaling exponents of indices and that of the $\epsilon$
parameter differ from each other at both solar minimum and maximum. This
difference between scaling exponents of $\delta(AU)$ and the driver
$\delta \epsilon$ is approximately the same at solar minimum and maximum.}
\item{A Fokker-Planck approach can be used to model the fluctuation PDF
for the geomagnetic indices in both phases of the solar cycle and the $\epsilon$
at solar maximum to a good approximation}
\end{enumerate}
The approximate statistical self-similarity found for the indices for solar
minimum and maximum and the $\epsilon$ at solar maximum is consistent with complex
multi-scale processes such as turbulence or Self-Organized Criticality (SOC).
The distinct values found for scaling exponents may reflect physical differences
in the solar wind and the magnetosphere but may also be due to the very different
way in which these quantities are derived. The fluctuations in the $AU$ index
depend on the solar cycle but the scaling exponent is distinct from that of
$\epsilon$ fluctuations.
Interestingly, the difference between scaling exponents of $\delta(AU)$ and the
driver $\delta \epsilon$ appears to be approximately constant. 
These observations, when combined together, suggest that the process
involved in generating fluctuations in the $AU$ index is coupled to the solar
wind driver, as seen in the solar cycle dependence. In contrast to the $AU$
index fluctuations, these in the $AL$ index are nearly insensitive to the
change in solar cycle implying that the $AL$ index is a measure of activity
intrinsic to the magnetosphere. This is consistent with the $AU$ index
more closely monitoring activity on the day-side and $AL$ reflecting activity
in the magnetotail.

The self-similar scaling of fluctuations allows us to model their statistics using
a Fokker-Planck approach. We obtained analytically a functional form of the
fluctuation PDF which approximates the measured PDF rather well. We stress that
such an approach links the statistical features discussed here to dynamical
modeling of the time series via stochastic Langevin equations.

\section{Acknowledgments} 
B.~Hnat acknowledges support from the PPARC, S.~C.~Chapman from the Radcliffe
Institute and G. Rowlands from the Leverhulme Trust.
We thank R.P Lepping and K. Ogilvie for provision of data from the NASA WIND
spacecraft, the ACE SWEPAM instrument team and the ACE Science Center for
providing the ACE data and the WDC for the geomagnetic indices data.

\end{document}